\documentclass[prd,twocolumn,showpacs,amsmath,amssymb,nofootinbib,superscriptaddress]{revtex4}
\usepackage{graphicx}
\usepackage[pdftex]{color}
\usepackage{bm}
\usepackage{amsfonts}
\usepackage{amssymb}
\usepackage{color}
\usepackage{float}
\usepackage{amsmath}
\usepackage{dcolumn}
\usepackage{hyperref}
\usepackage{amsfonts}

\def\nn{\nonumber}

\def\be{\begin{equation}}
\def\ee{\end{equation}}
\def\bea{\begin{eqnarray}}
\def\eea{\end{eqnarray}}
\def\eqi{\begin{equation}}
\def\eqf{\end{equation}}
\def\eqia{\begin{eqnarray}}
\def\eqfa{\end{eqnarray}}
\def\lcdm{$\Lambda$CDM }

\begin{document}

\title{Cosmological constraints and comparison of viable $f(R)$ models}

\author{Judit P\'erez-Romero}\email{judit.perezr@estudiante.uam.es}

\author{Savvas Nesseris}\email{savvas.nesseris@csic.es}
\affiliation{Instituto de F\'isica Te\'orica UAM-CSIC, Universidad Auton\'oma de Madrid, Cantoblanco, 28049 Madrid, Spain}

\date{\today}

\begin{abstract}
In this paper we present cosmological constraints on several well-known $f(R)$ models, but also on a new class of models that are variants of the Hu-Sawicki one of the form $f(R)=R-\frac{2\Lambda}{1+b\;y(R,\Lambda)}$, that interpolate between the cosmological constant model and a matter dominated universe for different values of the parameter $b$, which is usually expected to be small for viable models and which in practice measures the deviation from General Relativity. We use the latest growth rate, Cosmic Microwave Background, Baryon Acoustic Oscillations, Supernovae type Ia and Hubble parameter data to place stringent constraints on the models and to compare them to the cosmological constant model but also other viable $f(R)$ models such as the Starobinsky or the degenerate hypergeometric models. We find that these kinds of Hu-Sawicki variant parameterizations are in general compatible with the currently available data and can provide useful toy models to explore the available functional space of $f(R)$ models, something very useful with the current and upcoming surveys that will test deviations from General Relativity.
\end{abstract}

\maketitle

\section{Introduction}
Recent independent cosmological observations (see Ref.\cite{Ade:2015xua} and references therein) suggest that the  Universe is spatially flat and contains approximately $\sim 30\%$ of matter (dark and baryonic). The remaining $\sim 70\%$ is attributed to the so-called dark energy (DE) and is considered to be responsible for the accelerated expansion of the Universe, even though the mechanism behind such a phenomenon has not yet been directly identified.

Even though the true nature of the DE is elusive for the time being, in recent years a plethora of cosmological models that attempt to explain the accelerated expansion of the Universe have been created and can in general be split into two broad and general thematic categories. The first category retains General Relativity (GR) but contains new fields that have yet to be directly observed in Earth-based laboratories, see for example Refs.\cite{Copeland:2006wr,Caldwell:2009ix}. The second category is mainly based on covariant modifications of GR that make gravity weaker at large scales, hence explaining away the observed accelerated expansion of the Universe as a purely geometric effect \cite{Copeland:2006wr,Clifton:2011jh}.

While the observed kinds of matter cannot explain the accelerated expansion of the Universe as their equation-of-state (EoS) parameter $w$, which is defined as the ratio of the fluid's pressure to its density or $w=\frac{P}{\rho}$, is always positive since for baryons we have $w\simeq0$ and for radiation $w=\frac13$, some of the dark energy fluids mentioned earlier and the cosmological constant $\Lambda$ have an EoS of $w>-1$ and $w=-1$ respectively. However, in the case of modified gravity models the EoS parameter $w$ can enter into the phantom regime, namely $w<-1$, which would normally be attributed to ghost fields \cite{Nesseris:2006er}.

One of the most prominent cosmological models included in the second group is $f(R)$ gravity \cite{Starobinsky:1980te}. These models are simple extensions of the Einstein-Hilbert action, e.g. the model with Lagrangian of the type $f(R)=R+m R^2$ that was pioneered by Starobinsky as the first example of $f(R)$ models \cite{Starobinsky:1980te}, but at the same time they can be sufficiently broad that they incorporate interesting features of higher order gravity models. However, it has been shown \cite{Amendola:2006kh,Amendola:2006we} that most $f(R)$ models do not contain a proper matter era during the expansion history of the Universe and hence are not viable. A similar conclusion was also reached in Ref.~\cite{Amendola:2006we} using the autonomous systems approach of the $f(R)$, which provides interesting and unique insights on the evolution of the models. However it has been shown that matter domination can be achieved in a number of models, see Refs.~\cite{Nojiri:2006gh,Capozziello:2006dj,Odintsov:2017hbk,Elizalde:2010ts}.

More details on the properties of these models can be found in the reviews of Refs.~\cite{DeFelice:2010aj,Sotiriou:2008rp,Nojiri:2010wj,Nojiri:2017ncd} and specifically the details of the evolution of Newton's constant $G_{\textrm{eff}}$ and the density perturbations can be found in Refs.~\cite{Tsujikawa:2007gd,Tsujikawa:2007tg}, while somewhat older cosmological constraints on some of these theories can be found in Refs.~\cite{Basilakos:2013nfa,Abebe:2013zua,Dossett:2014oia,delaCruz-Dombriz:2015tye}, and some very interesting methods to reconstruct $f(R)$ models from observations can be seen in Refs.~\cite{Lee:2017lud,Lee:2017dox} .

Two very popular viable models with a proper matter era that also pass the solar system tests are the following:
\begin{enumerate}
\item
The Hu \& Sawicki (HS) model \cite{Hu:2007nk} with
\begin{equation}
\label{Hu}
f(R)=R-m^2 \frac{c_1 (R/m^2)^n}{1+c_2 (R/m^2)^n},
\end{equation}
\item
and the Starobinsky model \cite{Starobinsky:2007hu} with
\begin{equation}
\label{Star}
f(R)=R-c_1~m^2 \left[1-\left(1+R^2/m^{4}\right)^{-n}\right]\;,
\end{equation}
\end{enumerate}
where in both cases we have that $c_1$, $c_2$ are two free parameters, $m^2\simeq \Omega_{m0}H^{2}_{0}$ is of the order of the Ricci scalar $R_{0}$, $H_{0}$ is the Hubble constant, $\Omega_{m0}$ is the dimensionless matter density today; and $m$ and $n$ are positive constants with $n$ usually taking positive integer values ie $n=1, 2, \cdots$.

Even though originally it was claimed that these two models do not contain the cosmological constant $\Lambda$ at all, using simple algebraic manipulations one may show that both the Hu \& Sawicki and the Starobinsky models actually do contain $\Lambda$ and are simple extensions of the \lcdm model, where the deviation from \lcdm is characterized by a parameter $b$ as follows \cite{Basilakos:2013nfa,Bamba:2012qi}:
\begin{equation}
\label{mod-lcdm}
f(R)=R-2\Lambda\; \tilde{y}(R,b),
\end{equation}
where the function $\tilde{y}(R,b)$ is given by
\begin{equation}
\label{hsgr}
\tilde{y}(R,b)=1-\frac{1}{1+(R/b~\Lambda)^n},
\end{equation}
for the Hu \& Sawicki model with $\Lambda= \frac{m^2 c_1}{2c_2}$ and $b=\frac{2 c_2^{1-1/n}}{c_1}$, and by
\begin{equation}
\label{stargr}
\tilde{y}(R,b)=1- \frac{1}{\left(1+\left(\frac{R}{b \Lambda }\right)^2\right)^n},
\end{equation}
for the Starobinsky model where $\Lambda= \frac{c_1 m^2}{2}$ and $b=\frac{2}{c_1}$. As can be seen, both models have the same limits for $n>0$ \cite{Basilakos:2013nfa}:
\bea
\lim_{b\rightarrow0}f(R)&=&R-2\Lambda , \nn \\
\lim_{b\rightarrow \infty}f(R)&=&R,
\eea
and as a result both of the models reduce to \lcdm for $b\rightarrow 0$. Therefore, it is quite clear that both of the aforementioned models practically contain the cosmological constant $\Lambda$. This clearly demonstrates that the real reason for successfully passing all the observational tests is of course that they are small perturbations around the \lcdm model.

Furthermore, in Ref.~\cite{Basilakos:2013nfa} it was found that the parameter $b$ was of the order of $b\sim 0.1$ for the Hu \& Sawicki model thus making it practically indistinguishable from the \lcdm at the background level. This also has the interesting side-effect that, as it was shown in Ref.~\cite{Basilakos:2013nfa}, for small values of the parameter $b$ one is always able to find analytic approximations to the Hubble parameter for these models that work to a level of accuracy of better than $\sim10^{-5}\%$.

The goal of our paper is to systematically explore the functional space of the $f(R)$ theories that simultaneously are viable and at the same time behave as small perturbations around the \lcdm model. Without loss of generality we choose the general form of these $f(R)$ models to be of the type:
\be
\label{fRansatze}
f(R)=R-\frac{2\Lambda}{1+b~y(R,\Lambda)},
\ee
where $y(R,\Lambda)$ is a function of the Ricci scalar $R$ and of the cosmological constant $\Lambda$, while $b$ is a free parameter assumed to be small.

By choosing properly the function $y(R,b)$ one can always construct viable $f(R)$ models that behave as small perturbations around the \lcdm model but always contain the cosmological constant as a limiting case for $b\rightarrow 0$. The main advantage of the approach is that one can keep all the benefits of the \lcdm model, such as passing the solar system tests and having a proper matter era, but at the same time also exploring the available functional space of the $f(R)$ theories in a consistent and viable manner.

In the next sections we will test in-depth several ansatze for the function $y(R,\Lambda)$ and compare them to the Hu \& Sawicki, Starobinsky and other $f(R)$ models using the latest cosmological data, including the Joint Light-curve Analysis (JLA) type Ia supernovae, the Planck 2015 CMB shift parameters, the Baryon Acoustic Oscillation data and the ``Gold 2017" growth rate compilation presented in \cite{Nesseris:2017vor}.

The layout of our manuscript is as follows:
In Sec.~\ref{backevo} we discuss and review the general theory of the $f(R)$ models, in Sec.~\ref{funcforms} we present the collection of $f(R)$ lagrangians we will consider in our analysis, and in Sec.~\ref{constrs} we test and compare all the $f(R)$ lagrangians using the latest cosmological data. Finally, we present our conclusions in Sec.~\ref{conclusions}.

\section{The general theory \label{backevo}}

In this section we will briefly review the main theoretical framework of the $f(R)$ models. We will assume a spatially flat homogeneous and isotropic universe filled with nonrelativistic matter and radiation. Then, the modified Einstein-Hilbert action reads:

\be
S=\int d^{4}x\sqrt{-g}\left[  \frac{1}{2\kappa^{2}}f\left(  R\right)
+\mathcal{L}_{m}+\mathcal{L}_{r}\right],  \label{action1}%
\ee
where $\mathcal{L}_{m}$ is the Lagrangian of matter, $\mathcal{L}_{r}$ is the Lagrangian of radiation and
$\kappa^{2}=8\pi G_N$ is a constant where $G_N$ is the Newton's constant. Varying the action with respect to the metric, following the metric variational approach, we arrive at
\bea
F G_{\mu\nu}&-&\frac12(f(R)-R~F) g_{\mu\nu}+\left(g_{\mu\nu}\Box-\nabla_\mu\nabla_\nu\right)F\nn \\ &=&\kappa^{2}\,T_{\nu}^{\mu},
\label{EE}
\eea
where $F=f'(R)$, $G_{\mu\nu}$ is the Einstein tensor and $T_{\nu}^{\mu}$ is the energy-momentum tensor.

As mentioned, we will assume a spatially flat homogeneous and isotropic universe, hence we can use the Friedmann$-$Lemaitre$-$Robertson$-$Walker (FLRW) metric, which in Cartesian coordinates takes the familiar form:
\be
ds^{2}=-dt^{2}+a^{2}(t) d\vec{x}^{2} \label{metric},
\ee
where $a(t)=\frac{1}{1+z}$ is the scale factor and $z$ is the cosmological redshift.

In order to close the system of equations we also need the evolution equations for the density of the radiation and the cold dark matter. This can be done by modeling them as perfect fluids with $4-$velocity $U_{\mu}$ and an energy momentum tensor $T_{\nu}^{\mu}=(\rho+P)U^{\mu}U_{\nu}-P\,g_{\nu}^{\mu}$, where $\rho=\rho_{m}+\rho_{r}$ and $P=p_{m}+p_{r}$ are the total energy density and pressure of the fluid. In this case, $\rho_{m}$ is the matter density, $\rho_{r}$ corresponds to the density of the radiation and $p_{m}=0$, $p_{r}=\rho_{r}/3$ are the corresponding pressures. As is well known, the evolution equations for the matter and radiation come from the Bianchi identity $\nabla^{\mu}\,{T}_{\mu\nu}=0$ which in the context of the FLRW metric leads to the conservation laws:
\bea
\dot{\rho}_{m}&+&3H\rho_{m}=0, \\
\dot{\rho}_{r}&+&4H\rho_{r}=0, \label{frie3}%
\eea
where the dot corresponds to a derivative with respect to the cosmic time $t$ and $H\equiv\dot{a}/a$ is the Hubble parameter. The previous equations can now be solved to give the evolution of the density in terms of the scale factor as $\rho_{m}(a)=\rho_{m0}a^{-3}$ and $\rho_{r}(a)=\rho_{r0}a^{-4}$ respectively.

Using the FLRW metric of Eq.~(\ref{metric}) and the field equations (\ref{EE}) we can now derive the modified Friedmann's equations
\bea
3 F H^{2}&-&\frac{F R-f}{2}+3H\dot{F} =\kappa^{2}(\rho_{m}+\rho_{r}), \label{fried1}\\%
-2F\dot{H}&=&\kappa^{2}\left(\rho_{m}+\frac43\rho_{r}\right) +\ddot{F}-H\dot{F},\label{fried2}%
\eea
where $\dot{R}=aH\frac{dR}{da}$ and $F_{R}=\partial_R F=f''(R)$. Also, the Ricci scalar in this case is given by
\begin{equation}
R=g^{\mu\nu}R_{\mu\nu}= 6\left(  \frac{\ddot{a}}{a}+\frac{\dot
{a}^{2}}{a^{2}}\right)  =6(2H^{2}+\dot{H}) \;.\label{ricci}
\end{equation}

From Eqs.~(\ref{fried1}) and (\ref{fried2}) we can clearly see that it is impossible to solve them analytically in general, hence in what follows we will do so numerically when comparing the results of the $f(R)$ models with those of the \lcdm model. It is also quite interesting to derive the effective dark energy EoS parameter in terms of $E(a)=H(a)/H_{0}$
\begin{equation}
\label{eos222}
w(a)=\frac{-1-\frac{2}{3}a\frac{{d\rm lnE}}{da}}
{1-\Omega_{m}(a)},
\end{equation}
where
\be
\label{ddomm}
\Omega_{m}(a)=\frac{\Omega_{m0}a^{-3}}{E^{2}(a)} \;.
\ee
Clearly, in the case of the \lcdm model, which is described by $f(R)=R-2\Lambda$, the corresponding dark energy EoS parameter is exactly equal to $-1$ and the Hubble parameter is given by
\be
H_{\Lambda}(a)/H_0=\left( \Omega_{m0}a^{-3}+\Omega_{r0}a^{-4}+1-\Omega_{m0}-\Omega_{r0}\right)^{1/2} \;\;.\label{Hlcdm}
\ee

Finally, we should also stress that in the case of the $f(R)$ theories Newton's constant is always time and scale dependent and given in the quasistatic and subhorizon approximation by  \cite{Tsujikawa:2007gd,Tsujikawa:2007tg}:
\be
G_{\textrm{eff}}/G_N=\frac1{F}\frac{1+4\frac{k^2}{a^2}\frac{F_{,R}}{F}}{1+3\frac{k^2}{a^2}\frac{F_{,R}}{F}}\label{geff},
\ee
where $G_N$ is the bare Newton's constant and $k$ is the scale of the Fourier modes. It should also be noted that a more accurate approximation for $G_{\textrm{eff}}$ was found in Ref.~\cite{delaCruzDombriz:2008cp}, while useful statistical tools to detect possible deviations of $G_{\textrm{eff}}$ from unity were derived in \cite{Nesseris:2011pc}.

In the case of the \lcdm model the previous expression gives $G_{\textrm{eff}}/G_N=1$ as expected. There are also other stringent gravity constraints on $G_{\textrm{eff}}/G_N$, which are $G_{\textrm{eff}}>0$ and $G_{\textrm{eff}}/G_N=1.09\pm 0.2$, that are derived by demanding that the gravitons carry positive energy and that the Big Bang Nucleosynthesis is not affected. Furthermore, following our notation and definitions, we should have $G_{\textrm{eff}}(a=1)/G_N=1$. As a result, the evolution of the matter density perturbation $\delta=\frac{\delta\rho_m}{\rho_m}$ is given by \cite{Tsujikawa:2007gd}:
\be
\ddot{\delta}+2 H \dot{\delta}=4 \pi G_{\textrm{eff}} \delta \label{growthode}.
\ee

In order to compare with observations we actually require the quantity $f\sigma_8(a)=f(a)\cdot \sigma_8(a)$, which is defined in terms of the growth rate $f(a)=\frac{d \log \delta}{d \log a}$ and $\sigma_8(a)=\sigma_{8,0}\frac{\delta(a)}{\delta(1)}$. Then, we can equivalently write it as $f\sigma_8(a)=\sigma_{8,0} \frac{\delta'(a)}{\delta(1)}$ and this quantity can be directly compared to the observational data. For more details on the growth factor and the compilation of the relevant data we refer the interested reader to Ref.~\cite{Nesseris:2017vor}. In the following, we will use for simplicity the subscript 0 in $\sigma_{8,0}$ to denote the present day value.

\section{The $f(R)$ functional forms and the numerical approach\label{funcforms}}

In this section we will now present the functional forms of the $f(R)$ lagrangians used in our analysis. First, as mentioned we will use the well-known Hu \& Sawicki \cite{Hu:2007nk} model as expressed by Eq.~(\ref{Hu}) and  later on for simplicity but no loss of generality we will set $n=1$.

As discussed in the Introduction, after simple algebraic manipulations Eq.~(\ref{Hu}) can also be written as
\bea
\label{Hu1}
f(R)&=& R- \frac{m^2 c_1}{c_2}+\frac{m^2 c_1/c_2}{1+c_2 (R/m^2)^n} \nn\\
&=& R- 2\Lambda\left(1-\frac{1}{1+(R/(b~\Lambda)^n}\right) \nn \\
&=& R- \frac{2\Lambda }{1+\left(\frac{b \Lambda }{R}\right)^n},
\eea where $\Lambda= \frac{m^2 c_1}{2c_2}$ and $b=\frac{2 c_2^{1-1/n}}{c_1}$. In this form it is clear that this model can be arbitrarily close to $\Lambda$CDM, depending on the parameters $b$ and $n$.

Second, we will also consider the Starobinsky \cite{Starobinsky:2007hu} model given by Eq.~(\ref{Star}), which can also be rewritten as
\be
\label{Star1}
f(R)=R-2\Lambda  \left(1- \frac{1}{\left(1+\left(\frac{R}{b \Lambda }\right)^2\right)^n}\right),
\ee where $\Lambda= \frac{c_1 m^2}{2}$ and $b=\frac{2}{c_1}$. In this form it is clear that this model can also be arbitrarily close to $\Lambda$CDM.

\begin{table*}[t!]
\caption{The ``Gold-2017" compilation of $f\sigma_8(z)$ measurements from different surveys, compiled in Ref.~\cite{Nesseris:2017vor}. In the columns we show in ascending order with respect to redshift, the name and year of the survey that made the measurement, the redshift and value of $f\sigma_8(z)$ and the corresponding reference and fiducial cosmology. These datapoints are used in our analysis in the next sections.
\label{tab:fsigma8data}}
\begin{centering}
\begin{tabular}{ccccccc}
Index & Dataset & $z$ & $f\sigma_8(z)$ & Refs. & Year & Notes \\
\hline
1 & 6dFGS+SnIa & $0.02$ & $0.428\pm 0.0465$ & \cite{Huterer:2016uyq} & 2016 & $(\Omega_m,h,\sigma_8)=(0.3,0.683,0.8)$ \\

2 & SnIa+IRAS &0.02& $0.398 \pm 0.065$ &  \cite{Turnbull:2011ty},\cite{Hudson:2012gt} & 2011& $(\Omega_m,\Omega_K)=(0.3,0)$\\

3 & 2MASS &0.02& $0.314 \pm 0.048$ &  \cite{Davis:2010sw},\cite{Hudson:2012gt} & 2010& $(\Omega_m,\Omega_K)=(0.266,0)$ \\

4 & SDSS-veloc & $0.10$ & $0.370\pm 0.130$ & \cite{Feix:2015dla}  &2015 &$(\Omega_m,\Omega_K)=(0.3,0)$ \\

5 & SDSS-MGS & $0.15$ & $0.490\pm0.145$ & \cite{Howlett:2014opa} & 2014& $(\Omega_m,h,\sigma_8)=(0.31,0.67,0.83)$ \\

6 & 2dFGRS & $0.17$ & $0.510\pm 0.060$ & \cite{Song:2008qt}  & 2009& $(\Omega_m,\Omega_K)=(0.3,0)$ \\

7 & GAMA & $0.18$ & $0.360\pm 0.090$ & \cite{Blake:2013nif}  & 2013& $(\Omega_m,\Omega_K)=(0.27,0)$ \\

8 & GAMA & $0.38$ & $0.440\pm 0.060$ & \cite{Blake:2013nif}  & 2013& \\

9 &SDSS-LRG-200 & $0.25$ & $0.3512\pm 0.0583$ & \cite{Samushia:2011cs} & 2011& $(\Omega_m,\Omega_K)=(0.25,0)$  \\

10 &SDSS-LRG-200 & $0.37$ & $0.4602\pm 0.0378$ & \cite{Samushia:2011cs} & 2011& \\

11 &BOSS-LOWZ& $0.32$ & $0.384\pm 0.095$ & \cite{Sanchez:2013tga}  &2013 & $(\Omega_m,\Omega_K)=(0.274,0)$ \\

12 & SDSS-CMASS & $0.59$ & $0.488\pm 0.060$ & \cite{Chuang:2013wga} &2013& $\ \ (\Omega_m,h,\sigma_8)=(0.307115,0.6777,0.8288)$ \\

13 &WiggleZ & $0.44$ & $0.413\pm 0.080$ & \cite{Blake:2012pj} & 2012&$(\Omega_m,h)=(0.27,0.71)$ \\

14 &WiggleZ & $0.60$ & $0.390\pm 0.063$ & \cite{Blake:2012pj} & 2012& \\

15 &WiggleZ & $0.73$ & $0.437\pm 0.072$ & \cite{Blake:2012pj} & 2012 &\\

16 &Vipers PDR-2& $0.60$ & $0.550\pm 0.120$ & \cite{Pezzotta:2016gbo} & 2016& $(\Omega_m,\Omega_b)=(0.3,0.045)$ \\

17 &Vipers PDR-2& $0.86$ & $0.400\pm 0.110$ & \cite{Pezzotta:2016gbo} & 2016&\\

18 &FastSound& $1.40$ & $0.482\pm 0.116$ & \cite{Okumura:2015lvp}  & 2015& $(\Omega_m,\Omega_K)=(0.270,0)$\\
\end{tabular}\par\end{centering}
\end{table*}

Another interesting parametrization we will consider is of the form derived in \cite{Nesseris:2013fca}
\bea
f(R)&=& R-2\Lambda+b~H_0^2\left(\frac{\Lambda}{R-3\Lambda}\right)^{\alpha}\cdot \nn \\
&& {}_2F_1\left(\alpha,\frac{3}{2}+\alpha,\frac{13}{6}+2\alpha,\frac{\Lambda}{R-3\Lambda}\right),\nn\\
\label{eq.99}
\eea
where $\alpha=\frac{1}{12}(-7+\sqrt{73})$ and ${}_2F_1\left(a_1,a_2,a_3,z\right)$ is a hypergeometric function. This model can be derived by requiring that the expansion history, i.e. the Hubble parameter, be given by that of the $\Lambda$CDM model, and then using the modified Friedmann equations to find the corresponding $f(R)$ function \cite{Nesseris:2013fca}.

We now present explicitly the rest of the parameterizations of the function $y(R,\Lambda)$, that appears in the lagrangian of Eq.~(\ref{fRansatze}) given by
\be
f(R)=R-\frac{2\Lambda}{1+b~y(R,\Lambda)},
\ee
and that we will use for the numerical analysis of the next sections. The functional forms of the function $y(R,\Lambda)$ that we will consider are power law functions of the form $(\Lambda/R)^n$ with $n=\frac12, 1, \frac32$ with the $n=1$ case corresponding to the HS model, various functions of the ratio $\Lambda/R$ such as ArcTanh$(\Lambda/R)$, $\sin(\Lambda/R)$, Sinh$(\Lambda/R)$, $e^{\Lambda/R}$, Tanh$(\Lambda/R)$, ArcSin$(\Lambda/R)$, $\ln{(\Lambda/R)}$, Tan$(\Lambda/R)$, ArcTan$(\Lambda/R)$, Erf$(\Lambda/R)$ and ArcSinh$(\Lambda/R)$.

The main advantage of these parameterizations is that they interpolate between extreme values as the Ricci scalar varies from zero to infinity, thus sampling the available functional space. We also consider a simple polynomial expansion in terms of $\Lambda/R$ of the form $y(R,\Lambda)=b(\frac{\Lambda}{R})+c(\frac{\Lambda}{R})^2$ but also a Pad\'{e} Approximant $y(R,\Lambda)=\frac{b(\frac{\Lambda}{R})+c(\frac{\Lambda}{R})^2}{1+p(\frac{\Lambda}{R})+n(\frac{\Lambda}{R})^2}$.

Furthermore, most models must possess a chameleon mechanism in order to be compatible with solar system tests of gravity \cite{Brax:2008hh,Lombriser:2012nn}. However, our parameterizations are just variations of the Hu-Sawicki type around the $\Lambda$CDM model, so they naturally also contain a chameleon mechanism, thus they should be compatible with solar system tests of gravity as well.

We should note that for our analysis in the next sections we will use the most recent compilations of the type Ia supernovae known as JLA, the Baryon Acoustic Oscillations (BAO) data, the Cosmic Microwave Background (CMB) shift parameters based on Planck 2015 and the $H(z)$ data compilation as presented in Ref.~\cite{Basilakos:2016nyg}. We have also used the ``Gold-2017" growth-rate data compilation of Ref.~\cite{Nesseris:2017vor}, shown for completeness in Table \ref{tab:fsigma8data}. For the in-depth details of the analysis of the JLA, BAO, CMB and $H(z)$ data we refer the interested reader to Ref.~\cite{Basilakos:2016nyg}, while for that of the new growth-rate data we refer the reader to Ref.~\cite{Nesseris:2017vor}.

Finally, the analysis of the aforementioned data and parameterizations was performed via a Markov Chain Monte Carlo (MCMC) code developed by one of the authors\footnote{The codes used in the analysis are freely available at \url{http://members.ift.uam-csic.es/savvas.nesseris/}. }. The results of fitting the previous parameterizations are discussed in the next sections and the best-fit parameters and the corresponding errors for the parameters of the models are shown in Tables \ref{tab.6}, \ref{tab.7} and \ref{tab.8}.

\section{Constraints on $f(R)$ models \label{constrs}}
\subsection{Constraints from the growth rate data}

Now that we have presented the $f(R)$ models that we will use in the analysis, we can compare how well they fit the available data. However, since the main difference of these models will be at the perturbations level, we will first start with fitting them to the growth rate data and in the next subsection we will also use the rest of the data.

The procedure in this case consists of solving the differential equations for the modified Friedmann equation (\ref{fried1}) numerically with initial conditions that correspond to the \lcdm model at high redshifts. We do this as we expect that any deviations from the cosmological model will only occur at low redshifts and we have actually confirmed this for all of our models. For the corrections needed for the growth rate, namely the ratio of the Hubble parameter times the angular diameter distance $H(a)d_A(a)$ for the real to the fiducial model, now the numerator will be different for each model. Furthermore, note that now we use $G_{\textrm{eff}}$ from Eq.~(\ref{geff}), so for each model we have to compute it and check if it fulfills the local gravity constraints. Regarding also $G_{\textrm{eff}}$, we set $k=0.1h$Mpc$^{-1}\approx 300H_0$ following Ref.~\cite{Nesseris:2017vor}.

In Tables \ref{tab.6}, \ref{tab.7} and  \ref{tab.8} we show the best fit parameters for the models presented before. We split the results into different tables because the models use different numbers of parameters, so in Table \ref{tab.6} are the models which are variants of the Hu and Sawicki (HS) model. In Table \ref{tab.7} are the other models with different number of parameters and in Table \ref{tab.8} we show the best-fit $\chi^2$ and the values of the Information Criteria AIC and BIC.

\begin{table}[!t]
\centering
\caption{Best fit parameters for the parameterizations based of the function $y(R,\Lambda)$ that appear in the lagrangian of Eq.~(\ref{fRansatze}).}
\label{tab.6}
\begin{tabular}{| c | c | c | c |}
\hline
$y(R,\Lambda)$ Model & $\Omega_{m0}$ & $\sigma_8$ & $b$ \\
\hline
$\Lambda/R (HS)$ & 0.27$\pm$0.02 & 0.80$\pm$0.03 & 0.05$\pm$0.21 \\
$\sqrt{\Lambda/R}$ & 0.27$\pm$0.02 & 0.80$\pm$0.03 & 0.03$\pm$0.12 \\
$(\Lambda/R)^{3/2}$ & 0.207$\pm$0.019 & 0.89$\pm$0.03 & 0.4$\pm$1.0 \\
ArcTanh$(\Lambda/R)$ & 0.128$\pm$0.012 & 1.09$\pm$0.04 & 2$\pm$2 \\
$\sin(\Lambda/R)$ & 0.167$\pm$0.017 & 0.96$\pm$0.04 & 0.8$\pm$1.5 \\
Sinh$(\Lambda/R)$ & 0.187$\pm$0.018 & 0.93$\pm$0.04 & 0.7$\pm$1.4 \\
$e^{\Lambda/R}$ & 0.170$\pm$0.016 & 0.91$\pm$0.03 & 0.18$\pm$0.13 \\
Tanh$(\Lambda/R)$ & 0.202$\pm$0.018 & 0.89$\pm$0.03 & 0.3$\pm$0.8 \\
ArcSin$(\Lambda/R)$ & 0.206$\pm$0.018 & 0.89$\pm$0.03 & 0.1$\pm$0.5 \\
$\ln{(\Lambda/R)}$ & 0.208$\pm$0.018 & 0.89$\pm$0.03 & 0.01$\pm$0.04 \\
Tan$(\Lambda/R)$ & 0.26$\pm$0.02 & 0.81$\pm$0.03 & 0.04$\pm$0.21 \\
ArcTan$(\Lambda/R)$ & 0.26$\pm$0.02 & 0.81$\pm$0.03 & 0.05$\pm$0.17 \\
Erf$(\Lambda/R)$ & 0.26$\pm$0.02 & 0.81$\pm$0.03 & 0.04$\pm$0.20 \\
ArcSinh$(\Lambda/R)$ & 0.26$\pm$0.02 & 0.80$\pm$0.03 & 0.05$\pm$0.18 \\
\hline
\end{tabular}
\end{table}

\begin{table*}[t!]
\centering
\caption{Best fit parameters for different $f(R)$ models.\label{tab.7}}
\begin{tabular}{| c | c | c | c | c | c | c |}
\hline
$f(R)$ Model & $\Omega_{m0}$ & $\sigma_8$ & $b$ & $c$ & $p$ & $n$\\
\hline
Starobinsky $n=1$ & 0.200$\pm$0.017 & 0.91$\pm$0.04 & 0.1$\pm$0.7 & \_ & \_ & \_ \\
Hypergeometric & 0.206$\pm$0.018 & 0.89$\pm$0.03 & 0.05$\pm$0.02 & - & \_ & \_ \\
Polynomial & 0.107$\pm$0.011 & 1.12$\pm$0.04 & 2.3$\pm$1.0 & 0.0010$\pm$0.0035 & \_ & \_ \\
Pad\'{e} Approximant & 0.088$\pm$0.009 & 1.19$\pm$0.05 & 2.6$\pm$0.7 & 0.0010$\pm$0.0016 & 0.0011$\pm$0.0018 & 0.0010$\pm$0.0017 \\
\hline
\end{tabular}
\end{table*}

By inspecting Tables \ref{tab.6} and \ref{tab.7}, we can see that all of the models that have provided a theoretical prediction capable of producing a viable fit to the data have the form $\Lambda/R$. In most of the models we can see that the values for the $b$ parameters are perfectly compatible with 0, making all of these models a priori compatible with $\Lambda$CDM. Only three of them have a value for $b$ larger than the predicted error: $e^{\Lambda/R}$, the polynomial expansion and the Pad\'{e} approximant. For the value of $\Omega_{m0}$, most of the models predict $\Omega_{m0}\in [0.20-0.27]$, a range of values which is in fact smaller than the one coming from Planck15/$\Lambda$CDM. However, for some models this value is even smaller, reaching its minimum for the Pad\'{e} approximant. For the $\sigma_{8,0}$ parameter the variation range is even wider.

After we have analyzed the parameters obtained from the different fits, we can now evaluate which model fits the data best. In this case we work with models that use different numbers of parameters, so we need to compute the indicators Akaike Information Criterion (AIC) and Bayesian Information Criterion (BIC). The AIC is defined as AIC$=\chi^2_{min}+2k$ while the BIC is defined as BIC$=\chi^2_{min}+k \ln N$, where $k$ is the number of free parameters of the model and $N$ the number of datapoints used in the analysis.

In general small differences in the AIC are not necessarily significant and in order to test the ability of the models to reproduce the data, we have to investigate the differences $\Delta \mathrm{AIC}= \mathrm{AIC}_2-\mathrm{AIC}_1$ and $\Delta \mathrm{BIC}= \mathrm{BIC}_2-\mathrm{BIC}_1$, where the subindex 2 denotes the model with the largest indicator and the subindex 1 the one with the lowest. Clearly, the higher the value of $|\Delta \mathrm{AIC}|$, the higher the evidence against the model with higher value of AIC. Specifically, for the BIC a difference of 2 is considered as positive evidence, while 6 or more is strong evidence in favor of the model with the smaller value. Similarly, for the AIC a difference in the range between 0 and 2 means that the two models have more or less the same support from the data as the best one, for a difference in the range between 2 and 4 this support is considerably less for the model with the larger AIC, while for a difference $>10$ the model with the larger AIC is practically irrelevant \cite{Liddle:2004nh,Nesseris:2010pc}. However, this interpretation is based on the Jeffrey's scale, which should be interpreted with care \cite{Nesseris:2012cq}. For further details on the criteria and a comparison between them see Ref.~\cite{Liddle:2007fy}.

\begin{table}[t!]
\centering
\caption{Best fit values for all the models including $\Lambda$CDM. The models are ordered from the smallest $\chi^2$ to the largest. For the $\Delta$AIC and $\Delta$BIC the comparisons are made respect to $\Lambda$CDM, because it is the model with the lowest indicators.}
\label{tab.8}
\begin{tabular}{| c | c | c | c | c | c | c |}
\hline
Model & $\chi^2$ & $\chi^2/\nu$ & AIC & $\Delta$AIC & BIC & $\Delta$BIC \\
\hline
Pad\'{e} & 11.44 & 0.95 & 23.44 & 7.67 & 28.78 & 11.23 \\
Polynomial & 11.47 & 0.82 & 19.47 & 3.71 & 23.03 & 5.49 \\
ArcTanh & 11.49 & 0.77 & 17.49 & 1.72 & 20.16 & 2.61 \\
$\sin$ & 11.62 & 0.78 & 17.62 & 1.85 & 20.29 & 2.74 \\
Sinh & 11.69 & 0.78 & 17.69 & 1.93 & 20.37 & 2.82 \\
$e^{\Lambda/R}$ & 11.72 & 0.78 & 17.72 & 1.95 & 20.39 & 2.84 \\
$(\Lambda/R)^{3/2}$ & 11.74 & 0.78 & 17.74 & 1.97 & 20.41 & 2.86 \\
Tanh & 11.74 & 0.78 & 17.74 & 1.98 & 20.42 & 2.87 \\
Hypergeometric & 11.77 & 0.84 & 17.77 & 2.00 & 20.44 & 2.89 \\
$\Lambda$CDM & 11.77 & 0.74 & 15.77 & 0.00 & 17.55 & 0.00 \\
ArcSin & 11.77 & 0.79 & 17.77 & 2.00 & 20.44 & 2.89 \\
ln & 11.77 & 0.79 & 17.77 & 2.00 & 20.44 & 2.89 \\
Starobinsky & 11.81 & 0.79 & 17.81 & 2.04 & 20.48 & 2.93 \\
Tan & 12.16 & 0.81 & 18.16 & 2.39 & 20.83 & 3.28 \\
ArcTan & 12.17 & 0.81 & 18.17 & 2.40 & 20.84 & 3.29 \\
Erf & 12.18 & 0.81 & 18.18 & 2.41 & 20.85 & 3.30 \\
ArcSinh & 12.25 & 0.82 & 18.25 & 2.48 & 20.92 & 3.37 \\
$\sqrt{\Lambda/R}$ & 12.27 & 0.82 & 18.27 & 2.50 & 20.94 & 3.39 \\
$\Lambda/R (HS)$ & 12.28 & 0.82 & 18.28 & 2.51 & 20.95 & 3.40 \\
\hline
\end{tabular}
\end{table}

The values of the fit with the $\chi^2$ and the AIC/BIC indicators are shown in Table \ref{tab.8}. Inspecting Table \ref{tab.8} we see that the model that fits better the growth rate data, ie with the lowest $\chi^2$ at the minimum is the Pad\'{e} approximant, despite the values for $b$ and $\Omega_{m0}$ being in some tension with $\Lambda$CDM. Nevertheless, when computing the AIC/BIC indicators that take into account the number of parameters of each model, then the one that seems to fit the data better overall is $\Lambda$CDM, because it is the model with the lowest values for both indicators. Following the AIC criteria, the models which are totally compatible with $\Lambda$CDM are ArcTanh, Sin, Sinh, Exponential, $(\Lambda/R)^{3/2}$, Tanh, ArcSin and logarithmic. Two models have a mild tension with $\Lambda$CDM, which are the Polynomial expansion and the hypergeometric. According to the BIC criteria, the $\Lambda$CDM model is even more favored. In both cases, the model which presents the largest discrepancy is the Pad\'{e} approximant.

Since we want to do a proper comparison of the models, we are going to analyze three of them more deeply. We choose the original HS model with $n=1$, the Pad\'{e} approximant and the ArcTanh models. The models based on HS follow approximately the same behavior in terms of the expansion history, so they are compatible with each other. From Table \ref{tab.8}, the Pad\'{e} approximant seems a peculiar case, since the $\chi^2/\nu$ is the best but the number of parameters that it uses is penalized by the AIC and the BIC criteria. Finally, we also choose the ArcTanh model because according to the indicators it is the more compatible model with $\Lambda$CDM.

\begin{figure*}[!t]
	\centering	
	\includegraphics[width = 0.4\textwidth]{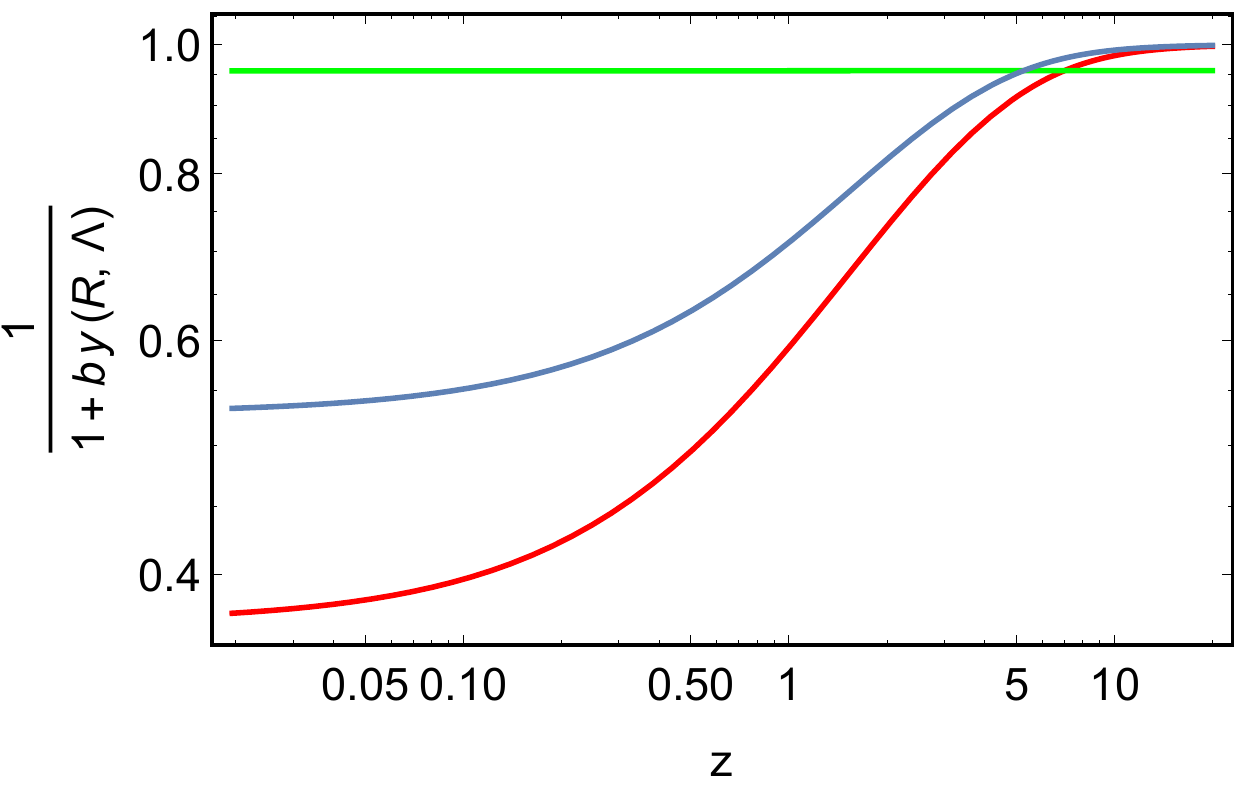}
    \includegraphics[width = 0.55\textwidth]{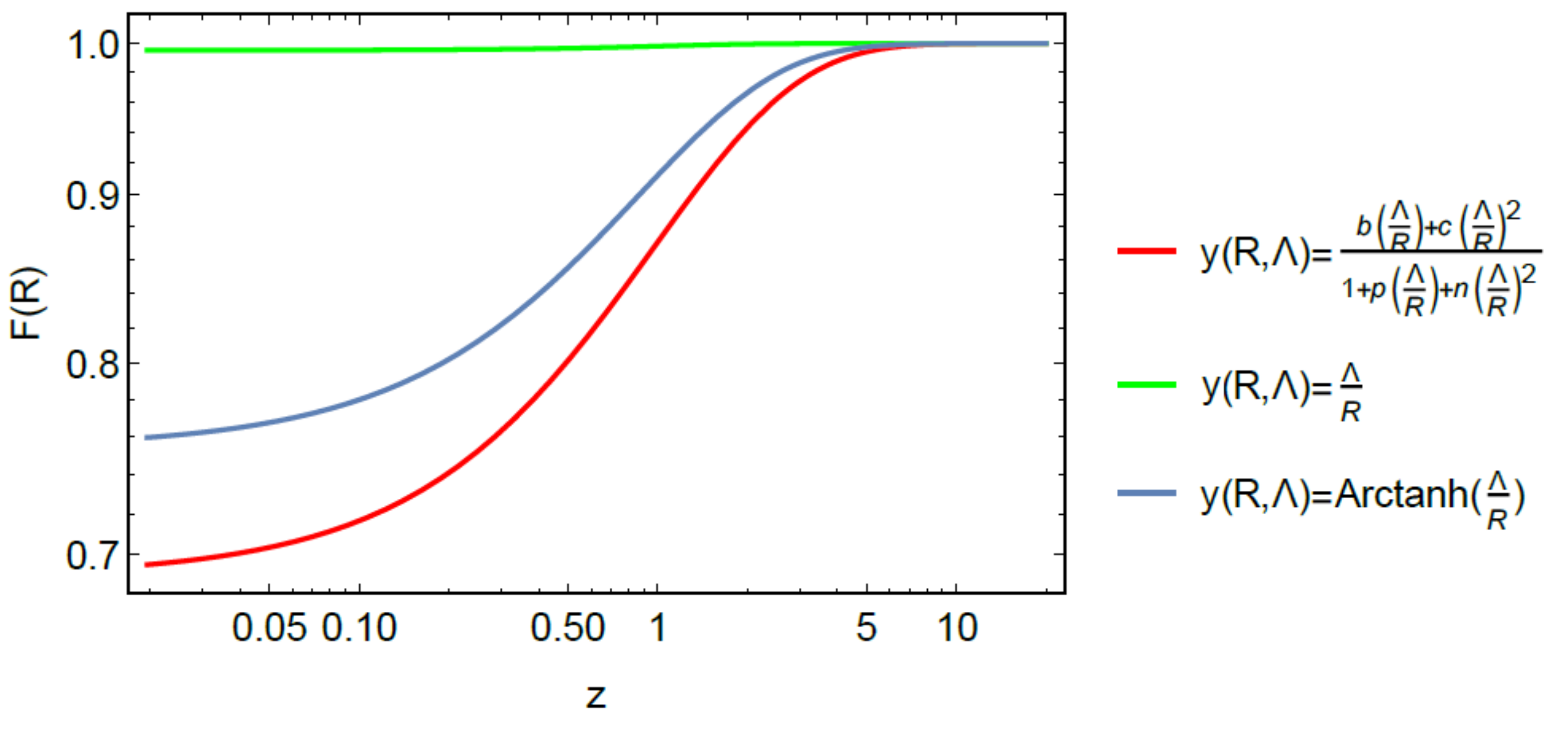}
	\caption{ \textbf{Left}: Plot of $\frac{1}{1+by(R,\Lambda)}$ as a function of the redshift for the three selected models. \textbf{Right}: Plot of $F(R)$ as a function of the redshift for the three selected models. Both of them are in the log-log scale. The values of the parameters used are the best fit in Tables \ref{tab.6} and \ref{tab.7}.}
	\label{fig.11}
\end{figure*}

The first comparison to check is the behavior of each $f(R)$ model in terms of the redshift. For this, we solve the modified Friedmann equation  of Eq.~(\ref{fried1}) and then implement the solution for $H(z)$ in the expression of the Ricci scalar Eq.~(\ref{ricci}). The plots are shown in  Fig.~\ref{fig.11}, where in the left panel  we show, according to the parametrization in Eq.~(\ref{fRansatze}), the correction to the $\Lambda$CDM model. In the case of $\Lambda$CDM, $\frac{1}{1+b~y(R,\Lambda)}$ should be 1. We see that the original HS model (green line) follows nearly the same behavior as $\Lambda$CDM, while the other two models interpolate between two different values. In both cases, the transition region is around $z\sim 0.5-1$, where the matter-DE equality happens. The Pad\'{e} approximant and the ArcTanh model reduce the weight of the $\Lambda$ term with respect to the $\Lambda$CDM model for small redshifts. This effect is more pronounced in the case of the Pad\'{e} approximant.

Although a priori one might think that this kind of parametrization would not describe properly the data, looking at the best fit values from Table~\ref{tab.8} we can see that both fit the data better than the HS model. The right panel of Fig.~\ref{fig.11} shows the derivative of $f(R)$ with respect to the Ricci scalar $F(R)=f'(R)$ for each model. For the $\Lambda$CDM model $F(R)=1$, so any deviation from unity shows deviation from the \lcdm model as well. In this plot we see the same behavior of the functions that we have commented on for the left panel: while the original HS model mimics quite well the $\Lambda$CDM behavior, the Pad\'{e} approximant and the ArcTanh parametrization decrease the ``effective" value of the cosmological constant for small redshifts with respect to the $\Lambda$CDM model.

\begin{figure*}[!t]
	\centering	
	\includegraphics[width = 0.45\textwidth]{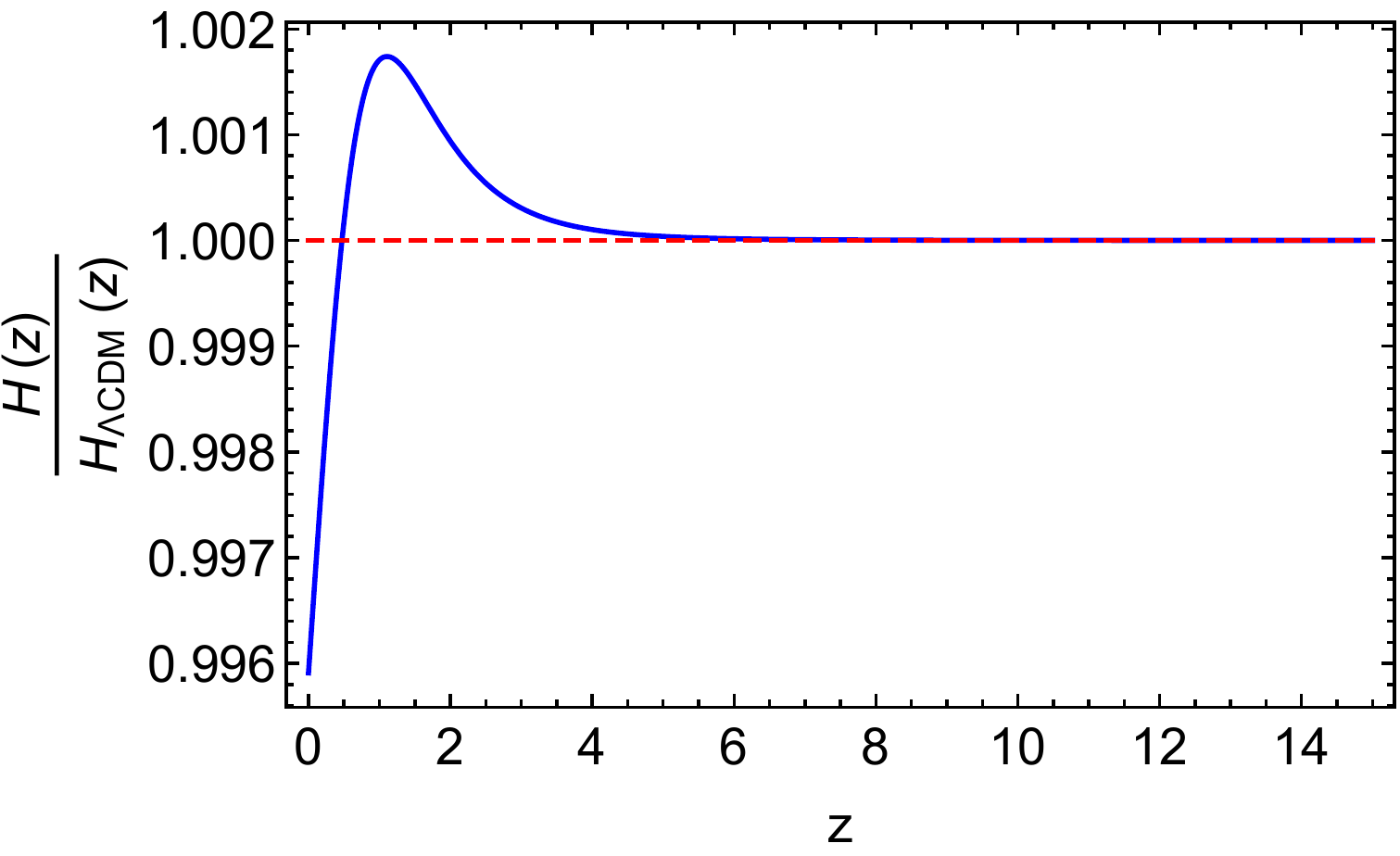}
    \includegraphics[width = 0.45\textwidth]{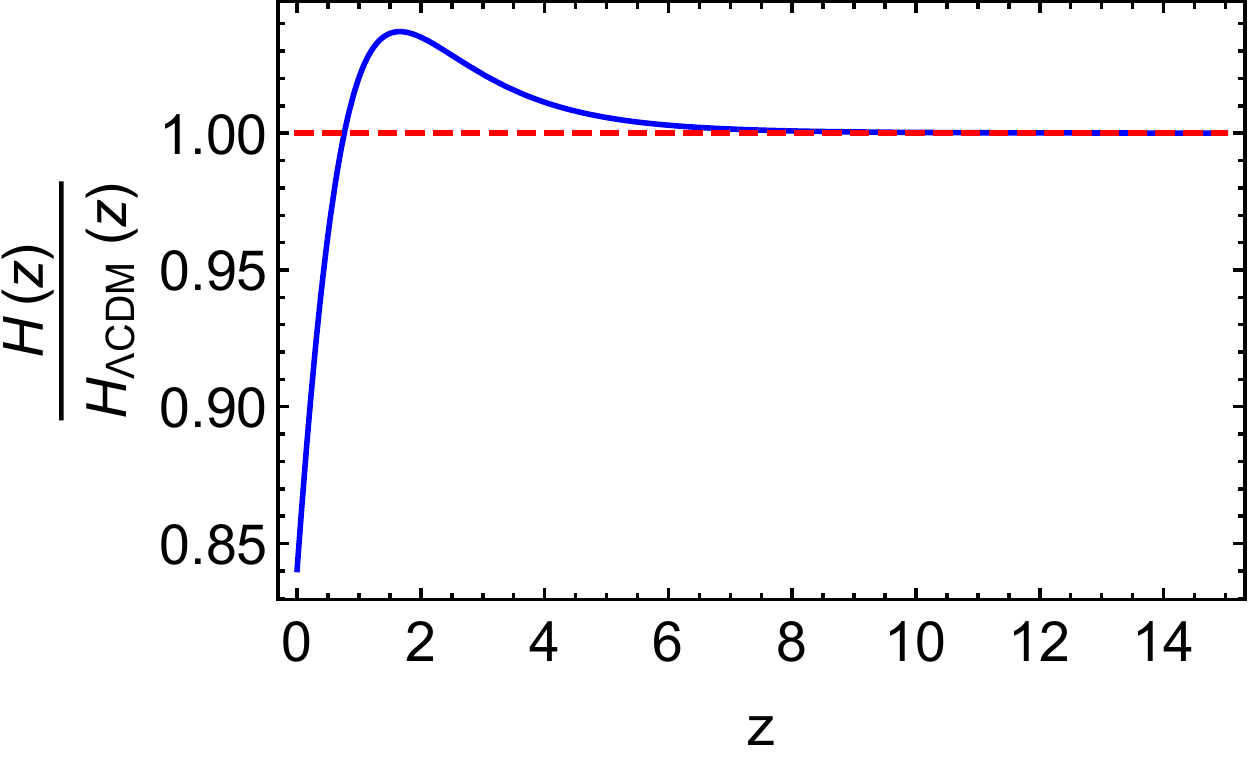}
    \includegraphics[width = 0.45\textwidth]{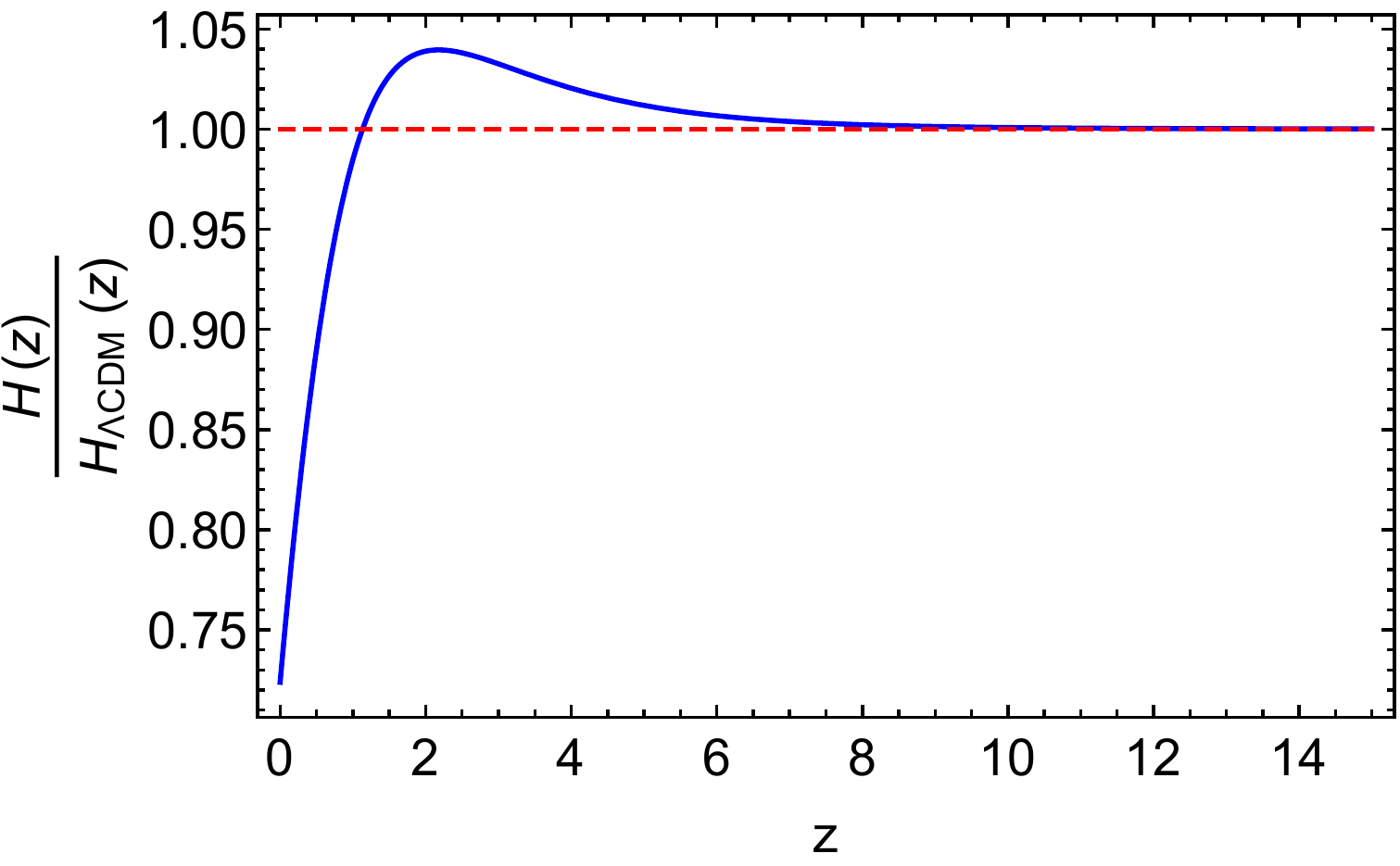}
	\caption{Comparison of the background expansion $H(z)$ in $f(R)$ gravity for the HS (top left), ArcTanh (top-right) and Pad\'{e} approximant (bottom) models to $\Lambda$CDM. The values of the parameters used are the best fit in Tables \ref{tab.6} and \ref{tab.7}.}
	\label{fig.12}
\end{figure*}

The second way we can compare the models with $\Lambda$CDM is through the background expansion of the Universe, $H(z)$. As the $\Lambda$CDM model uses the original Friedmann equations, we expect to see some differences arising from the modified Friedmann equation that we solve in the case of $f(R)$ gravity. For each model, we compare the result from solving Eq.~(\ref{fried1}) to the Hubble parameter of $\Lambda$CDM given by Eq.~(\ref{Hlcdm}). These plots are shown in Fig.~\ref{fig.12}.

The behavior of the $H(z)$ in the three cases is very similar: for high redshifts all of them recover $\Lambda$CDM (up to $z\sim 4-6$), for intermediate redshifts ($z\sim 2-4$) the Hubble parameter is larger than that for $\Lambda$CDM and for small redshifts ($z\sim 0-2$) the Hubble parameter decreases until it is smaller than that for $\Lambda$CDM. The fact that for large redshifts all of them follow $\Lambda$CDM is in agreement with our expectation that any deviations only appear at small redshifts. This new kind of behavior suggests that before entering in the DE domination era, the Universe should expand slightly faster than the prediction of the $\Lambda$CDM, until reaching a maximum and then expanding slower than for $\Lambda$CDM. This maximum does not coincide in any of the cases with the matter-DE equality. For the HS model we can see that the maximum difference between $H(z)$ is 0.4\%, 16\% in the case of the ArcTanh model and 28\% for the Pad\'{e} approximant. Having reached this point of the analysis, we can start thinking that the most different model from $\Lambda$CDM is the Pad\'{e} approximant, as the AIC and BIC indicators predicted.

\begin{figure}[!t]
	\centering	
	\includegraphics[width = 0.5\textwidth]{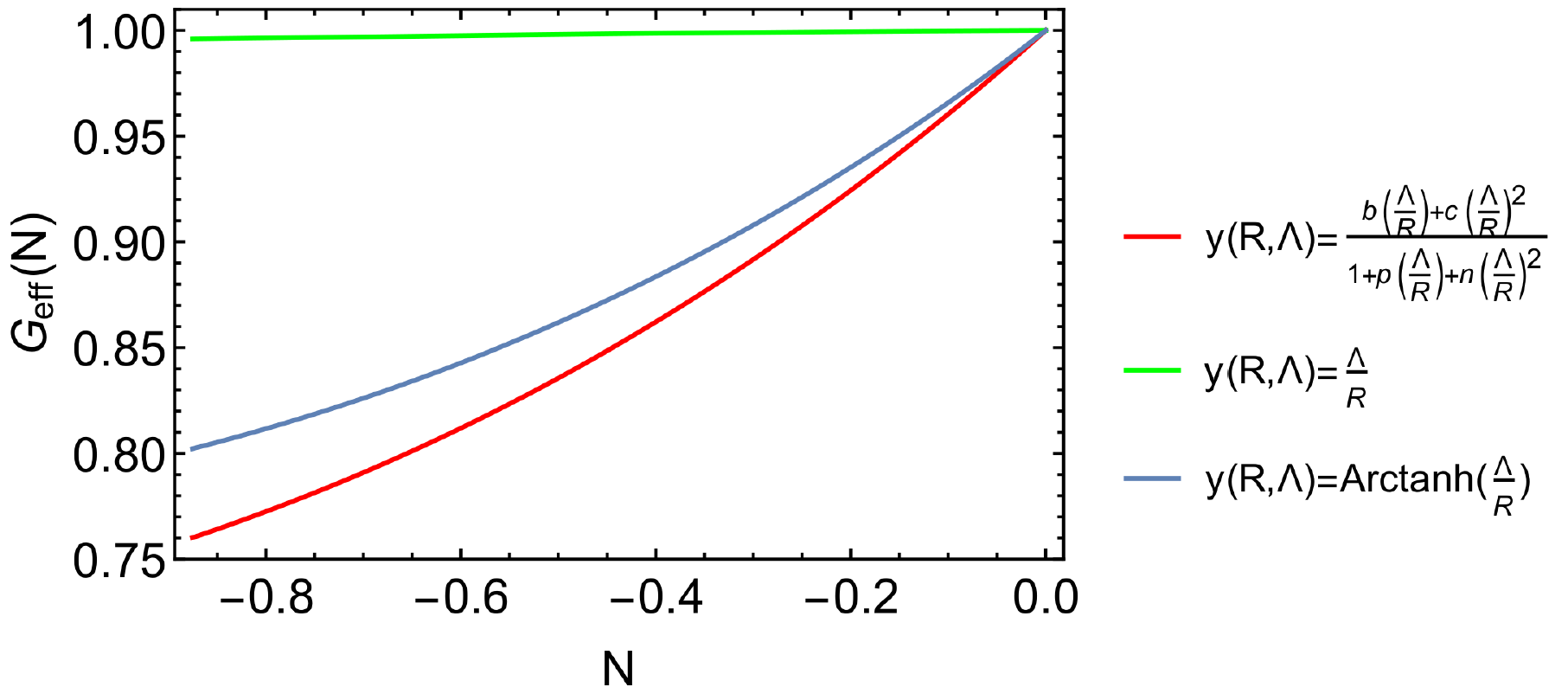}
	\caption{$G_{\textrm{eff}}/G_N$ as a function of the number of e-folds $N$. The values of the parameters used are the best fit in Tables \ref{tab.6} and \ref{tab.7}.}
	\label{fig.13}
\end{figure}

In Fig.~\ref{fig.13} we can see the effective Newton's constant $G_{\textrm{eff}}$ for each model as a function of the number of e-folds $N=\ln a$. This plot is really necessary since the models have to fulfill the local gravity constraints in order to be viable. The plot extends from today ($N=0$), until the maximum redshift that the ``Gold 2017'' provides, which is $z=1.4$. Although the $G_{\textrm{eff}}$ function is evaluated in the same range as $H(z)$, only the range that we have fitted is reliable. From this figure we can see again that the model that is more similar to $\Lambda$CDM is the original HS while the one with the highest deviation $\sim 25\%$ is the Pad\'{e} approximant. As we can only rely on this plot until the redshift of the data, we can say that all of them fulfill the conditions $G_{\textrm{eff}}>0$ and $G_{\textrm{eff}}(N=0)/G=1$, but the ArcTanh model and the Pad\'{e} approximant have a nonzero derivative so they would be in some tension with the local gravity constraints. As mentioned, our data only go up to $z\simeq1.4$, so the constraint from the Big Bang Nucleosynthesis (BBN) cannot be checked.

Regarding the local gravity constraints we should mention that while all models should pass them in order to be viable, it not straight-forward to include this information directly in our analysis as another datapoint. The reason is that Newton's constant, even though it is a very important quantity, is currently only indirectly inferred from other data, so we cannot add it directly in the likelihood as another datapoint or prior, eg in the form $\chi^2=\left(\frac{G_N -G_{\textrm{eff}}}{\sigma_{G_N}}\right)^2$. On the other hand, $G_{\textrm{eff}}$ also appears in our analysis via the differential equation for the growth rate given by Eq.~(\ref{growthode}), thus it at least enters the analysis from there.

\begin{figure*}[!t]
	\centering	
	\includegraphics[width = 0.45\textwidth]{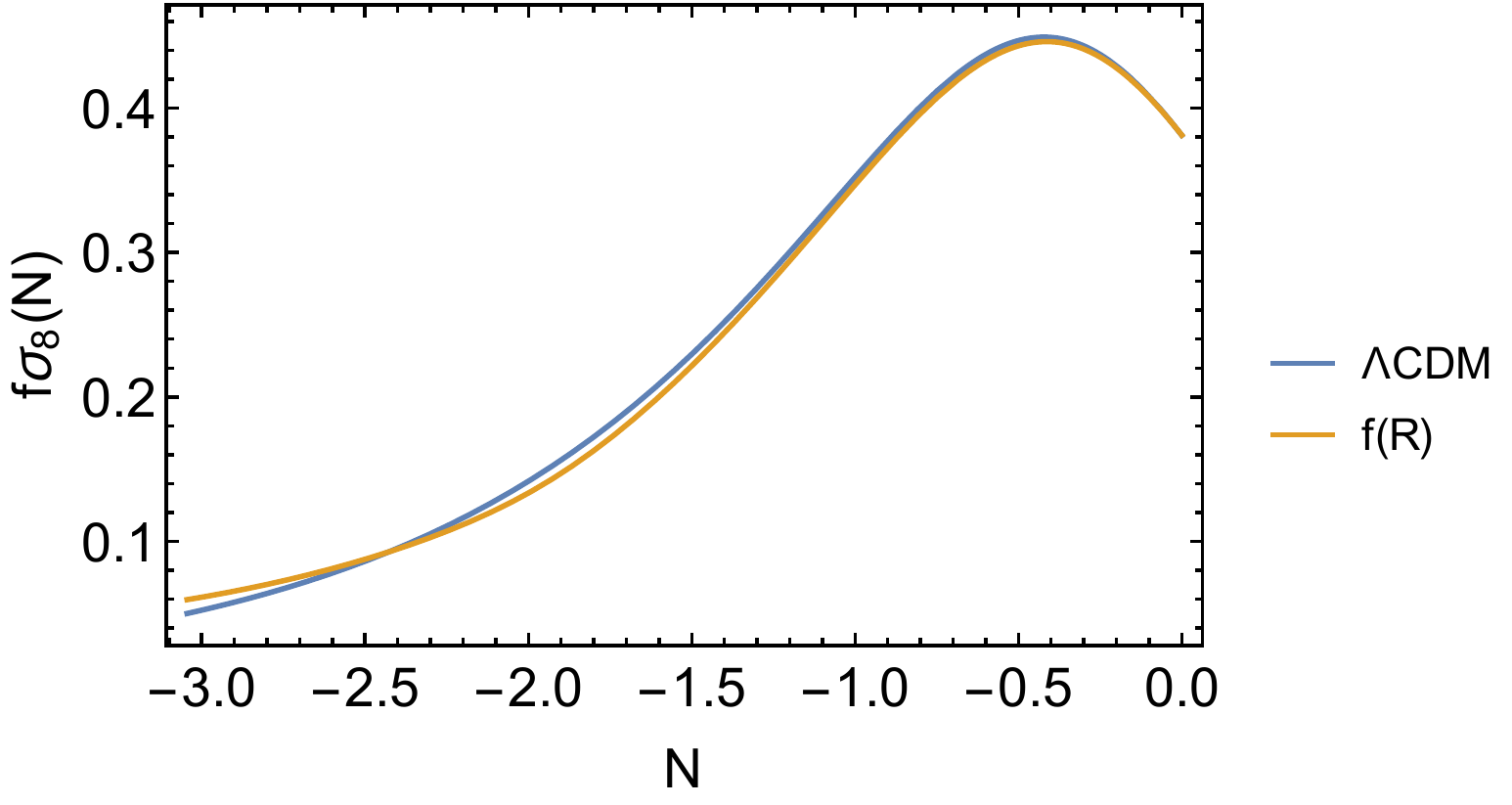}
    \includegraphics[width = 0.45\textwidth]{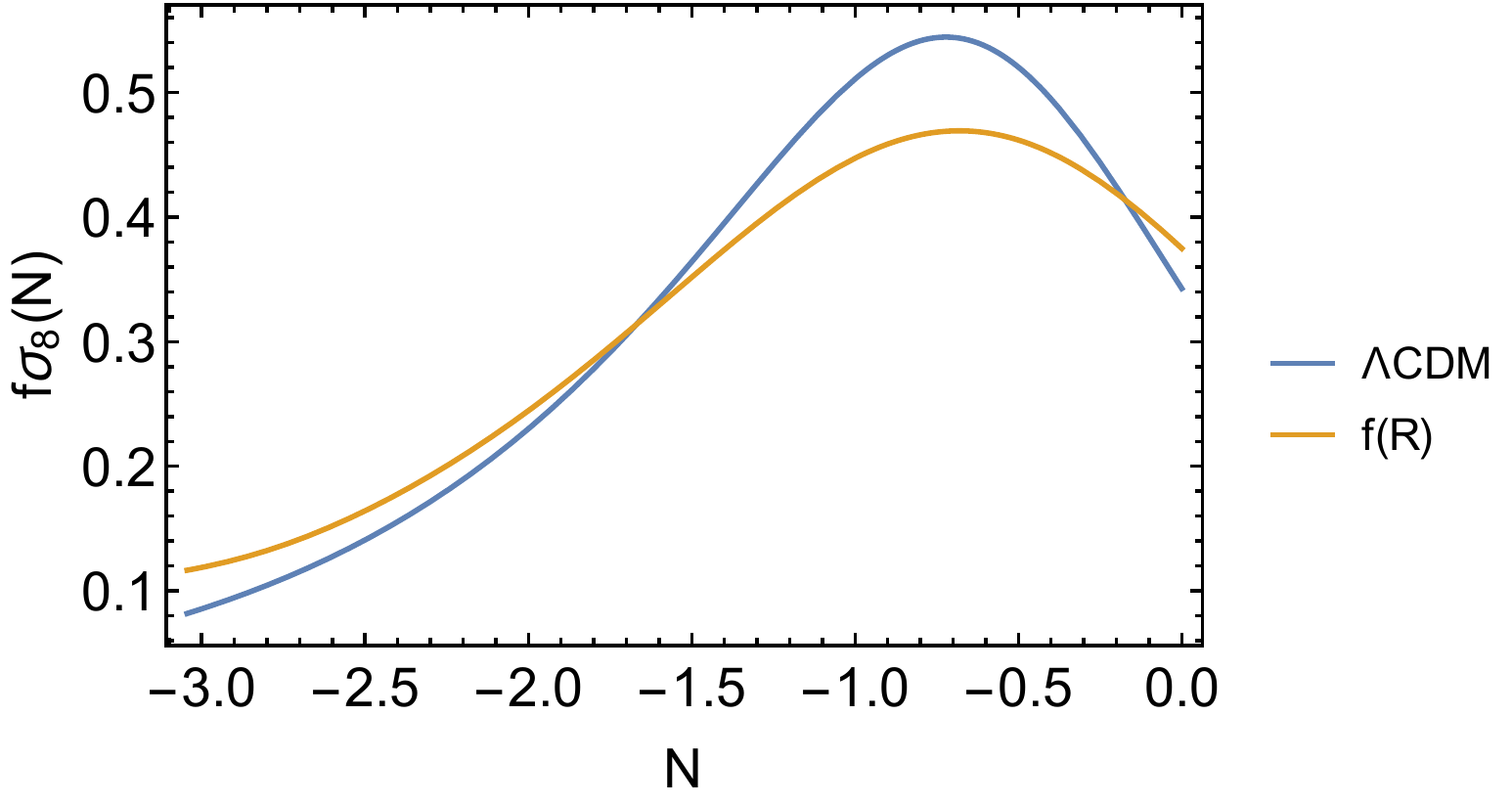}
    \includegraphics[width = 0.45\textwidth]{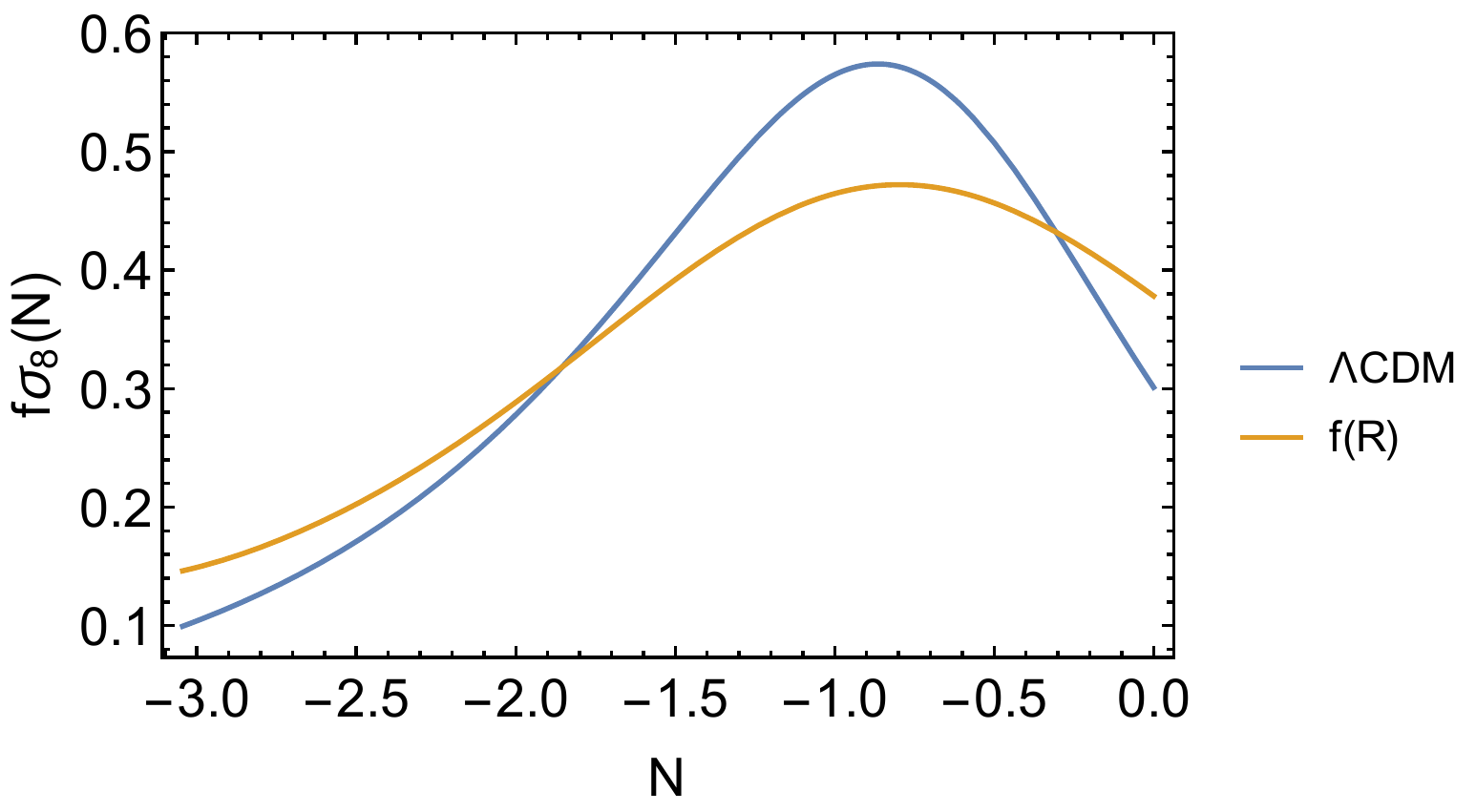}
	\caption{The theoretical predictions for $f\sigma_8$ for the HS (top left), ArcTanh (top-right) and Pad\'{e} approximant (bottom) models as a function of the number of e-folds $N$ (yellow line) compared to the prediction from $\Lambda$CDM (blue line). The values of the parameters used are shown in Tables \ref{tab.6} and \ref{tab.7}.}
	\label{fig.14}
\end{figure*}

Once we have constrained the background expansion and the $G_{\textrm{eff}}$ we are able to compare the different predictions for the growth rate $f\sigma_8$. This comparison is shown in Fig.~\ref{fig.14}. As we have already seen in the previous comparisons, the original HS model follows closely the same growth of structure as $\Lambda$CDM. For the ArcTanh and the Pad\'{e} models we can see that although the overall behavior is the same the growth of structures decreases with respect to $\Lambda$CDM in the matter era. This kind of feature is in line with the observations for low to intermediate redshifts that show a lack of structure, meaning a lack of gravitational power \cite{Kunz:2015oqa}. Then we can see that for these kind of observations, such as the ones for the ``Gold 2017'' dataset, the $f(R)$ models are a possibly viable alternative gravity theory to take into account. This lack of gravitational power can be related to the background expansion of the Universe, since for the same redshifts we observe a faster expansion and a lack of growth of structures. A side note is that the $\Lambda$CDM prediction in each case is different because we have used for each plot the values for the parameters from the best fit parameters.

\begin{figure*}[!t]
	\centering	
	\includegraphics[width = 0.4\textwidth]{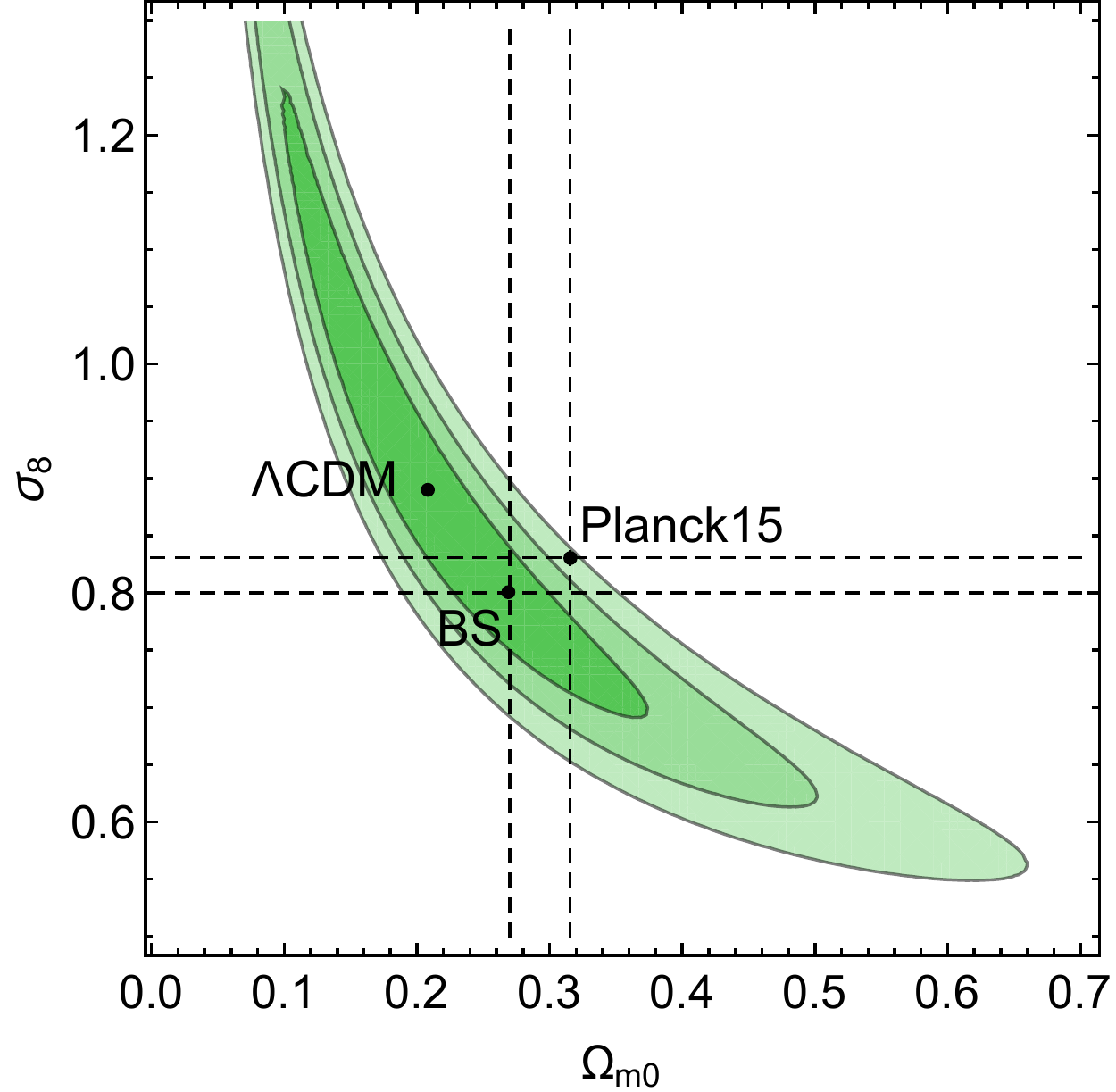}
    \includegraphics[width = 0.4\textwidth]{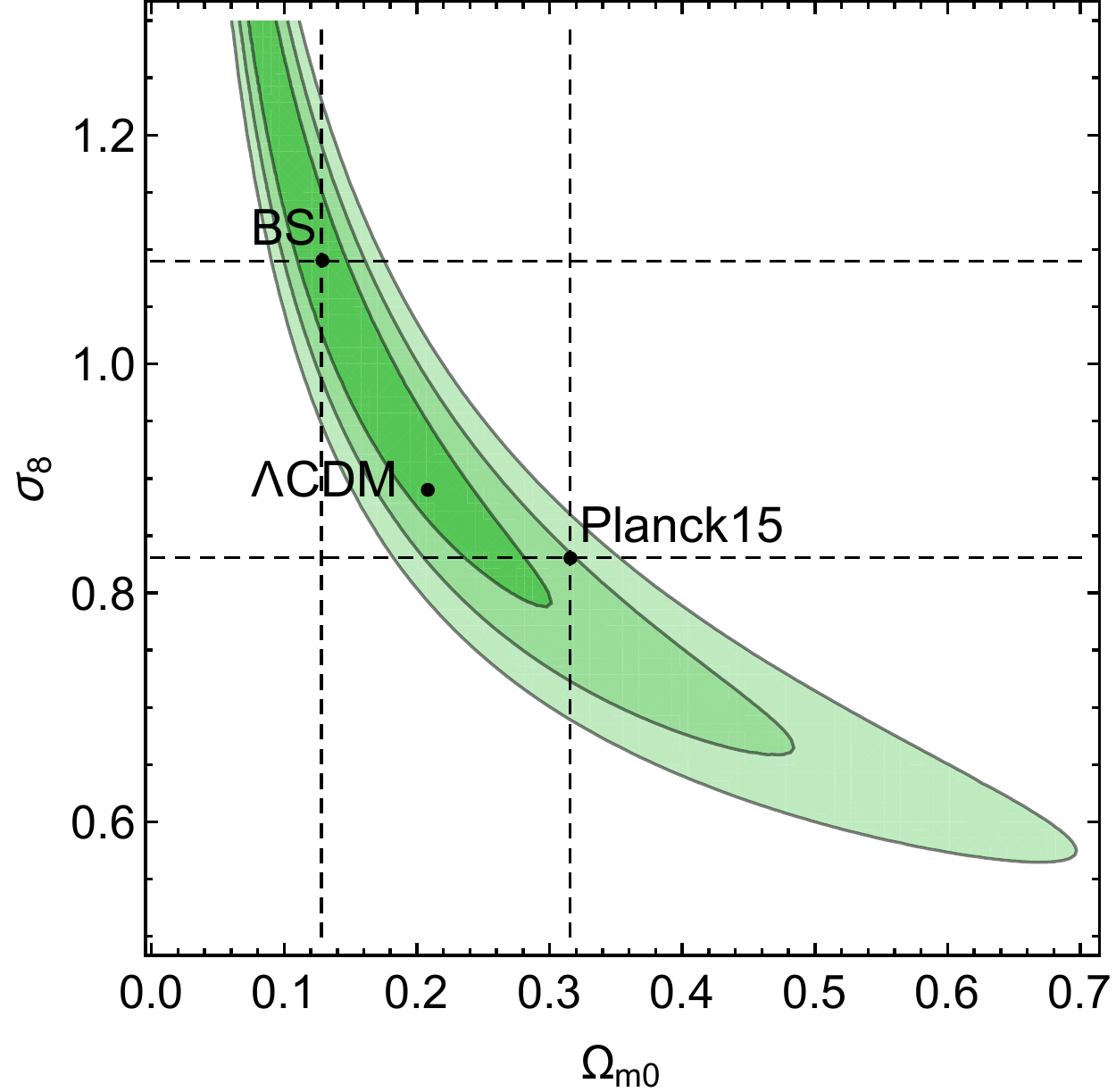}
    \includegraphics[width = 0.4\textwidth]{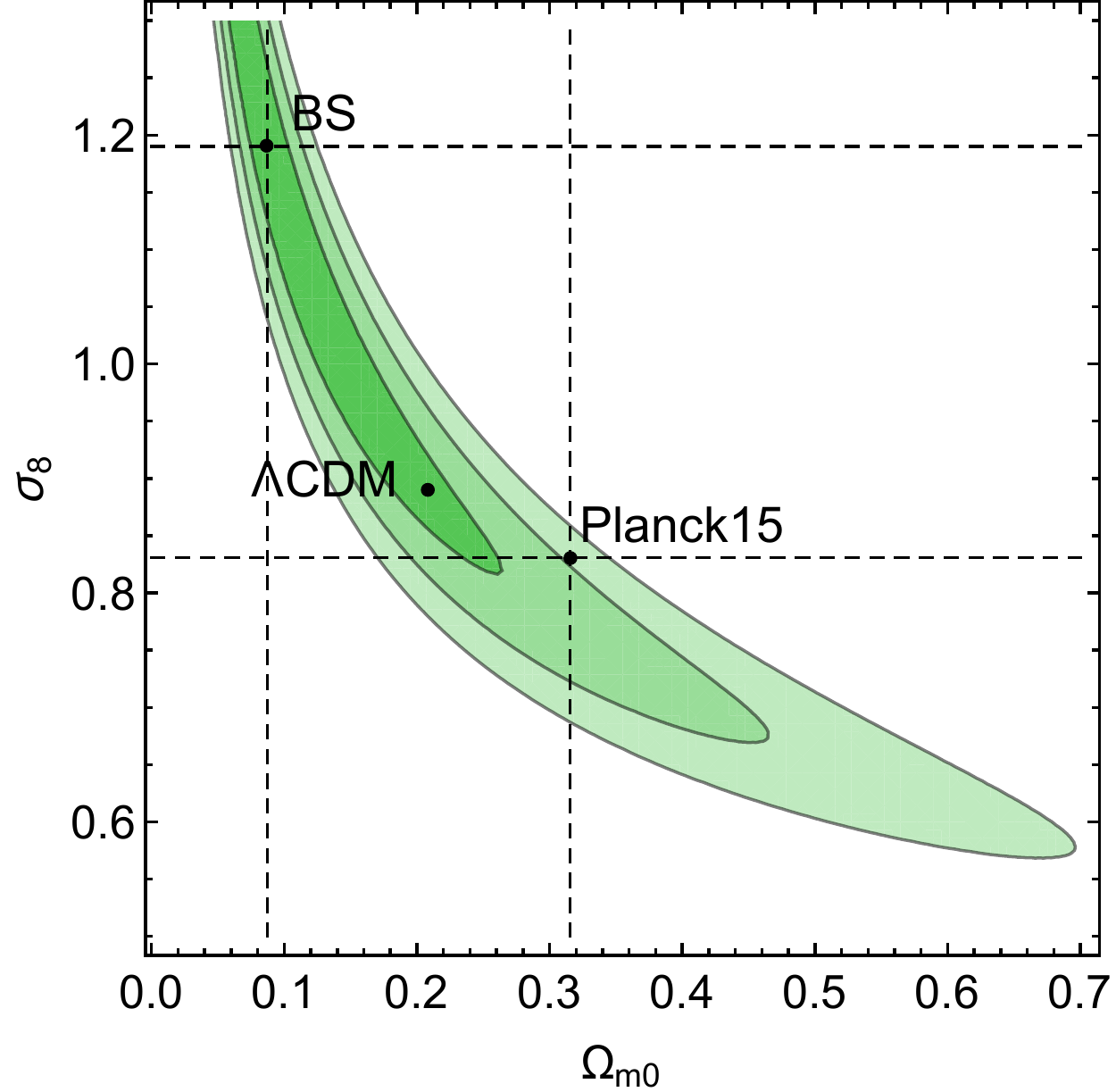}
	\caption{The 68.3\%, 95.4\% and 99.73\% confidence contours for the best fit parameters (BS) of the HS $n=1$ model (top left), the ArcTanh model (top right) and a Pad\'{e} approximant (bottom). We present also the best fit from Planck15/$\Lambda$CDM and from $\Lambda$CDM to the ``Gold 2017'' growth rate data.}
	\label{fig.15}
\end{figure*}

We also want to constrain the values of the best fit parameters for each model using as a reference the parameters from Planck15/$\Lambda$CDM. For this, we use again the confidence level regions in the plane $(\Omega_{m0},\sigma_{8,0})$. We show a contour for each model in Fig.~\ref{fig.15}.

We can see that for the three contours the values obtained for the three models are totally compatible with the value obtained from the fit of the ``Gold 2017'' growth rate dataset to $\Lambda$CDM. Furthermore, we can still see the previous tendency of the models. While the HS model obtains values which are closer to the ones obtained from $\Lambda$CDM and in the case of the ArcTanh and the Pad\'{e} approximant we get values which at first glance may seem incompatible.

However, looking at the contour we see that they are less than 1$\sigma$ away. If we compare the best fit values of each model with the Planck15/$\Lambda$CDM values instead, we find that the closest parameters are the ones from the ArcTanh model, being less than 2$\sigma$ away. However, when using our new parametrizations, in all cases the tension is eased as the best fit for the $\Lambda$CDM model when using the growth data only, is within $1\sigma$ from the $f(R)$ best-fit. Therefore, for these models the existing tension that was found for $\Lambda$CDM in Ref.~\cite{Nesseris:2017vor} is then alleviated.

The main conclusion of our analysis is that we cannot only take into account the results from the fits measured by the merit functions; in this case we use the $\chi^2$ or the indicators AIC or BIC. For a model to be cosmologically viable not only does it have to fit the data well, but it also has to fulfil the local gravity conditions for the $G_{\textrm{eff}}$. From the models that we have worked with, we have seen that even though the HS does not have the best $\chi^2$, it is the most similar model to $\Lambda$CDM when all the features are taken into account. The Pad\'{e} approximant is very penalized for the large number of parameters in the AIC and BIC indicators and also all the comparisons regarding the $\Lambda$CDM model conclude that it is the most different one, but not to the point of being totally excluded.

Finally, it should be noted that the ArcTanh model shows the smallest difference between the AIC and BIC compared to $\Lambda$CDM but with a significant first derivative at $z=0$. Also, the different comparisons that we perform show that its behavior is in between the HS model and the Pad\'{e} approximant. The final result that we obtain is that this model is also capable of releasing some of the existent tension between the growth rate data best fit and Planck/$\Lambda$CDM, as well as the previous parametrization of $G_{\textrm{eff}}$, and of opening the possibility to search for new physics in the direction of $f(R)$ gravity.

\subsection{Constraints from all of the data}

\begin{table*}[!t!]
\centering
\caption{Best fit parameters for a selection of $f(R)$ models using all the data. In the last case, the ${}_2F_1$ model corresponds to the degenerate $f(R)$ model of Ref.~\cite{Nesseris:2013fca} given by Eq.~(\ref{eq.99}). The values for the $\chi^2_{min}$ and the information criteria parameters are given in Table~\ref{tab.all1}\label{tab.all}}
\begin{tabular}{ccccccccc}
\hline
Model & $\alpha$ & $\beta$&$\Omega_{m0}$ & $\Omega_{b}h^2$ & $h$ & $\sigma_8$ & $b$ \\
\hline
\lcdm  & $0.141\pm0.003$ & $3.103\pm0.007$ &$0.315\pm0.004$ & $0.02224\pm0.00010$ & $0.673\pm0.003$& $0.743\pm0.029$ & $-$ \\
ArcTanh & $0.141\pm0.005$ & $3.099\pm0.008$ &$0.314\pm0.003$ & $0.02226\pm0.00011$ & $0.674\pm0.002$& $0.750\pm0.024$ & $0.010 \pm0.007$ \\
HS &$0.141\pm0.006$ & $3.097\pm0.009$ &$0.317\pm0.007$ & $0.02221\pm0.00012$ & $0.672\pm0.005$& $0.747\pm0.030$ & $0.010 \pm0.004$ \\
$\ln$ &$0.141\pm0.005$ & $3.099\pm0.008$ &$0.313\pm0.004$ & $0.02230\pm0.00011$ & $0.673\pm0.003$& $0.748\pm0.035$ & $0.005 \pm0.001$ \\
$\sin$ &$0.141\pm0.005$ & $3.096\pm0.009$ &$0.316\pm0.002$ & $0.02223\pm0.00011$ & $0.672\pm0.002$& $0.744\pm0.028$ & $0.050 \pm0.010$ \\
${}_2F_1$ & $0.141\pm0.006$ & $3.098\pm0.002$ &$0.317\pm0.004$ & $0.02222\pm0.00011$ & $0.672\pm0.003$& $1.023\pm0.032$ & $5.706 \pm0.343$ \\
\hline
\end{tabular}
\end{table*}

\begin{table}[!t!]
\centering
\caption{The values of the information criteria parameters for the $f(R)$ models of Table~\ref{tab.all}. In the second column we indicate the number of free parameters $q$, while in all cases the number of datapoints is $N=806$.\label{tab.all1}}
\begin{tabular}{ccccccc}
\hline
Model & $q$ &$\chi^2_{min}$& AIC & $\Delta$AIC& BIC & $\Delta$BIC\\
\hline
\lcdm   & 6  & $744.737$& $756.737$& $0.000$& $784.890$& $0.000$\\
arcTanh & 7  & $744.893$& $758.893$& $2.156$& $791.738$& $6.848$\\
HS      & 7  & $744.912$& $758.912$& $2.175$& $791.757$& $6.867$\\
$\ln$   & 7  & $745.108$& $759.108$& $2.371$& $791.953$& $7.063$\\
$\sin$  & 7  & $744.774$& $758.774$& $2.037$& $791.619$& $6.729$\\
${}_2F_1$ & 7 & $744.037$& $758.037$& $1.300$& $790.882$& $5.992$\\
\hline
\end{tabular}
\end{table}

In this section we will now consider only some of the models of the previous section and fit them in addition to the growth data, also to the most recent compilations of the type Ia supernovae known as JLA, the Baryon Acoustic Oscillations (BAO) data, the Cosmic Microwave Background (CMB) shift parameters based on Planck 2015 and the $H(z)$ data compilation as presented in Ref.~\cite{Basilakos:2016nyg}.

We note that in this case, all of the models have the same six parameters ($\alpha$, $\beta$, $\Omega_{m0}$ , $\Omega_{b}h^2$ , $h$ , $\sigma_{8,0}$), but in addition the $f(R)$ models also have the parameter $b$, thus totaling seven parameters, while the \lcdm model has only six. Furthermore, the number of datapoints now is $N=3+9+740+18+36=806$ for the CMB shift parameters, BAO, supernovae JLA, growth rate and $H(z)$ data respectively. Using a MCMC approach as before, we find the best-fit parameters of the models and subsequently the values for the Information Criteria. The former are shown in Table \ref{tab.all} and the latter in Table \ref{tab.all1}.

As can be seen, in all cases of the Hu \& Sawicki variant models, the best-fit value of the ``perturbation" parameter $b$ is indeed quite small and much smaller than 1, however in the case of the degenerate Hypergeometric model of Ref.~\cite{Nesseris:2013fca} given by Eq.~(\ref{eq.99}), the parameter is much larger, but this is a distinct case with regard to the rest. Surprisingly, as can be seen Table \ref{tab.all1} in this case the degenerate model also has the best fit to the data with $\delta \chi^2 =\chi^2_{\Lambda CDM}-\chi^2_{{}_2F_1}\sim 0.7$ with respect to the \lcdm model, but also a value for $\sigma_8$ which is significantly larger. The reason for this is that while it has the same background expansion history as the \lcdm model, it can further optimize the fit of the growth data, thus reducing the total $\chi^2$ but at the cost of the higher value for $\sigma_{8,0}$.

Furthermore, we note that all the HS variant models have a value for $\sigma_{8,0}$ which is compatible with that of \lcdm and of the Planck 15 data but is in contrast to the much higher values from the growth data alone. This tension is well known (see eg Ref.~\cite{Nesseris:2017vor}), and even though one may naively expect to resolving it by using a modified gravity model to resolve it, in this case we find that the fit and the best-fit value for $\Omega_{m0}$ are dominated by the Planck data. This is also in agreement with the results found in Ref.~\cite{Basilakos:2017rgc}.

Regarding the rest of the models, the ArcTanh and HS models have a $\chi^2$ very close to that of the \lcdm model with a difference of only $\delta \chi^2=\chi^2_{\Lambda CDM}-\chi^2_{arcTanh} \simeq -0.156$ and $\delta \chi^2=\chi^2_{\Lambda CDM}-\chi^2_{HS} \simeq -0.175$, however the differences in the AIC and BIC criteria are significant. In both cases the difference of the AIC is larger than 2 indicating some support for the \lcdm model. Similarly, for the BIC the difference is larger than 6 indicating strong support for the \lcdm model. An exception to this is the degenerate model which is on par with the \lcdm model.

\section{Conclusions \label{conclusions}}

In our present study we have considered a plethora of widely used but also new $f(R)$ models and compared them by using the growth rate data ``Gold 2017" compilation of Ref.~\cite{Nesseris:2017vor}, but also the most recent compilations of the type Ia supernovae known as JLA, the Baryon Acoustic Oscillations (BAO) data, the Cosmic Microwave Background (CMB) shift parameters based on Planck 2015 and the $H(z)$ data compilation as presented in Ref.~\cite{Basilakos:2016nyg}.

The familiar $f(R)$ models include the Hu \& Sawicki and Starobinsky models given by Eqs.~(\ref{Hu}) and (\ref{Star}) respectively, but also the degenerate hypergeometric model of Ref.~\cite{Nesseris:2013fca} given by Eq.~(\ref{eq.99}) that has the same expansion history as \lcdm but different perturbations and an evolving Newton's constant. Our new models included variants of the HS model given by Eq.~(\ref{fRansatze}) where for the function $y(R,\Lambda)$ we chose a set of functional forms, given in Table~\ref{tab.8}, but also a Pad\'{e} approximant and a polynomial expansion.

We found that most of the models can successfully fit the current growth data on their own and we used this as an opportunity to further study the desired properties of models that do so. As seen in Figs.~\ref{fig.11} and \ref{fig.12}, the best-fitting and most successful models deviate strongly from \lcdm at low redshifts, but still have values for the AIC and BIC information criteria that are comparable to the \lcdm model.

Then, we chose a subset of the available models and fitted them to all of the data at our disposal [the growth data, the SnIa, the CMB, the BAO and $H(z)$]. In this case we found that most of the $f(R)$ models are in some tension with \lcdm as evidenced with the increased values of the AIC and BIC in Table \ref{tab.all1}, except for the degenerate hypergeometric model that shares the same expansion history as \lcdm but has different perturbations.

Using our analysis, not only did we place stringent constraints on a plethora of old and new $f(R)$ models using the most recent cosmological data, but we also characterized the properties of any $f(R)$ model that could be viable and pass all the required conditions as discussed in the previous sections. Furthermore, as can be seen in Fig.~\ref{fig.15}, where we show the confidence contours for the parameters $\Omega_m$ and $\sigma_8$, the Planck15 \lcdm best fit is in all cases $3 \sigma$ away from the best fit \lcdm using the growth data only.

However, when using our new parametrizations, in all cases the tension is eased as the best fit \lcdm using the growth data only is within $1\sigma$ from the $f(R)$ best-fit and as a result, using the alternative cosmologies the tension between the CMB and growth-rate data is alleviated.

Finally, we note that a parametrization for the $f(R)$ lagrangian of the form of Eq.~(\ref{fRansatze}), could in principle be a useful tool in constructing viable models that are variants of the HS model and interpolate between a \lcdm and matter dominated model, that is also in good agreement with the observations. In principle, one could also reconstruct an appropriate lagrangian following a top-bottom approach, ie have all the desired properties in mind and then find the lagrangian itself. Models such as that could be useful with the current, future and up-coming surveys that aim to test deviations from Einstein's theory of General Relativity in the coming years.

\section*{Acknowledgements}
The authors acknowledge support from the Research Project FPA2015-68048-03-3P [MINECO-FEDER] and the Centro de Excelencia Severo Ochoa Program SEV-2016-0597. S.~N. also acknowledges support from the Ram\'{o}n y Cajal program through Grant No. RYC-2014-15843.

\raggedleft
\bibliography{bibliography}
\end{document}